\documentclass[conference,a4paper]{IEEEtran}
\IEEEoverridecommandlockouts
\usepackage{cite}
\usepackage{amsmath,amssymb,amsfonts}
\usepackage{booktabs}
\usepackage{algorithmic}
\usepackage{graphicx}
\usepackage{textcomp}
\usepackage{xcolor}
\def\BibTeX{{\rm B\kern-.05em{\sc i\kern-.025em b}\kern-.08em
    T\kern-.1667em\lower.7ex\hbox{E}\kern-.125emX}}
\begin{document}

\title{Low-Data Predictive Maintenance of Railway Station Doors and Elevators Using Bayesian Proxy Flow Modeling
\thanks{Funded by the European Union. Views and opinion expressed are however those of the author(s) only and do not necessarily reflect those of the European Union or Europe’s Rail Joint Undertaking. Neither the European Union nor the granting authority can be held responsible for them. The project FP3-IAM4Rail is supported by the Europe's Rail Joint Undertaking and its members.}
}

\author{
\IEEEauthorblockN{Waldemar Bauer}
\IEEEauthorblockA{
\textit{Department of Automatic Control \& Robotics}\\
\textit{AGH University of Krakow}\\
Krakow, Poland \\
0000-0002-8543-0995\\
bauer@agh.edu.pl}
\and
\IEEEauthorblockN{Jerzy Baranowski}
\IEEEauthorblockA{
\textit{Department of Automatic Control \& Robotics}\\
\textit{AGH University of Krakow}\\
Krakow, Poland \\
0000-0003-3313-581X\\
jb@agh.edu.pl}
}

\maketitle

\begin{abstract}
This paper proposes a low-data predictive maintenance framework for automatic doors and elevators in a railway station building. The method is intended for assets without direct condition monitoring, where only aggregate passenger traffic information and expert knowledge about movement patterns are available. Passenger flows are modeled on a reduced station graph using a Bayesian formulation with uncertain totals and routing shares. The inferred flows are converted into approximate operating-cycle loads for doors and elevators through simple stochastic proxy relations. These loads are combined with uncertain age- and cycle-based maintenance thresholds to estimate the probability that predefined maintenance conditions have been reached. A cost-aware scheduling model is then used to align maintenance activities while accounting for service costs, disruption, delay penalties, and grouping opportunities within each asset class. The framework is illustrated on a simulated case study reflecting a real station layout. The results show that proxy operational data can support maintenance scheduling with low incremental implementation cost and can improve alignment relative to a calendar-based policy.
\end{abstract}


\section{Introduction}

Predictive maintenance (PdM) and condition-based maintenance (CBM) have been extensively studied in machinery diagnostics, prognostics, and maintenance decision support. Foundational review papers describe CBM as a workflow linking data acquisition, feature extraction, diagnostics, prognostics, and maintenance decisions, while more recent surveys emphasize that practical PdM deployments remain strongly constrained by uncertainty, data quality, and implementation cost \cite{Jardine2006CBM,Lei2018Prognostics,Ahmad2012CBMDecision,Nunes2023PdMChallenges}. In many reported applications, however, the implicit assumption is that maintenance-relevant signals are available from dedicated sensors or supervisory systems. This assumption is natural for high-value or safety-critical installations, but it is much less realistic for numerous widely used devices whose individual instrumentation cost is difficult to justify.

This limitation is particularly visible in public-service and building-infrastructure assets such as automatic doors, elevators, escalators, and access systems. Although these devices are operationally important, they are often maintained using fixed schedules, inspections, and service routines rather than direct monitoring of technical condition. As a result, there is a mismatch between the promise of PdM and the data actually available in practice. In such settings, maintenance decisions must be supported using indirect, uncertain, and often aggregated operational information rather than dedicated condition-monitoring measurements. This broader asymmetry between maintenance need and data availability is discussed in \cite{BaranowskiSymmetrySubmitted}, which motivates the present work from a general low-data PdM perspective.

Bayesian approaches are especially relevant in this context because they allow maintenance planning under sparse data, uncertain parameters, and expert-informed prior assumptions. Earlier work has shown the usefulness of Bayesian formulations for preventive maintenance modeling, reliability-oriented decision support, and maintenance scheduling when observations are limited \cite{Percy1996BayesianPM,Percy1997SparseDataPM,Percy2002BayesianReliability,Bousquet2015BayesianGamma}. This makes Bayesian modeling a natural choice for low-data PdM settings, where uncertainty is not a secondary issue but a defining characteristic of the problem. The same argument also underlies the project-level research perspective developed in \cite{BaranowskiSymmetrySubmitted}, where low direct observability is treated as a structural feature of many economically relevant maintenance problems rather than as an exceptional case.

In this paper, we consider such a setting for automatic doors and elevators in a railway station building. The proposed approach does not rely on direct sensing of degradation or operating cycles. Instead, it uses proxy operational information that may already be available in practice, namely aggregate passenger traffic estimates and expert-defined average routing proportions over a reduced station graph. This choice is motivated by the observation that railway-station movement can be represented at an aggregate level without resorting to full microscopic pedestrian tracking, and that graph-based station models can support demand and usage inference from indirect information sources \cite{Haenseler2016TrainStationFacilities,Haenseler2017TrainStationOD}. From the maintenance side, recent railway-oriented work also indicates the importance of integrating predictive information with scheduling and operational priorities \cite{Gerum2019RailScheduling,Consilvio2024RailPriority}.

The main idea is to construct a probabilistic chain from passenger demand to maintenance scheduling. First, passenger flows associated with embarking and disembarking travelers are inferred on a reduced graph of the station building using a Bayesian model with uncertain totals and routing shares. Next, the inferred flows are translated into approximate cycle loads for automatic doors and elevators through simple stochastic proxy relations reflecting their different operating mechanisms. These load estimates are then combined with uncertain age- and cycle-based maintenance thresholds in order to estimate the probability that predefined maintenance conditions have been reached. Finally, these probabilities are used in a cost-aware scheduling model that aligns maintenance activities in time while accounting for service cost, disruption, delay penalties, and grouping opportunities within each asset class.

The contribution of the paper is threefold. First, a Bayesian proxy-flow model is formulated for a reduced railway-station graph in a low-data setting where only aggregate traffic information and expert-defined routing shares are available. Second, the paper introduces device-specific proxy load estimators for doors and elevators, allowing uncertain passenger flows to be converted into maintenance-relevant usage measures. Third, a probabilistic scheduling framework is proposed for aligning predefined maintenance activities under uncertainty while exploiting grouping effects within homogeneous asset classes.

The proposed framework is illustrated on a simulated case study reflecting the layout and circulation logic of the recently renovated \L{}\'od\'z Kaliska railway station while avoiding disclosure of proprietary operational data used in the broader project context. The intention is not to replace direct condition monitoring where such monitoring is available and economically justified, but to demonstrate that useful predictive maintenance support can also be constructed for unsensored assets using already available proxy information.

The remainder of the paper is organized as follows. Section~\ref{sec:method} presents the proposed framework, including the passenger-flow model, device load estimation, and maintenance-condition formulation. Section~\ref{sec:scheduling} introduces the maintenance scheduling model. Section~\ref{sec:simulation} describes the simulated case study, while Section~\ref{sec:results} presents the illustrative results. Final conclusions are given in Section~\ref{sec:conclusion}.

\section{Proposed Framework}
\label{sec:method}
The proposed framework consists of three linked inference layers. First, aggregate passenger demand is propagated through a reduced station graph in order to estimate uncertain flows associated with maintained assets. Second, these flows are converted into proxy operating loads for automatic doors and elevators using device-specific stochastic relations. Third, the resulting usage estimates are combined with uncertain time- and cycle-based maintenance thresholds in order to quantify the probability that a predefined maintenance condition has been reached. This probability is then used in the scheduling layer presented in the following section.
\subsection{Passenger Flow Model}

The station is represented by a reduced directed graph \(G=(V,E)\), where nodes denote relevant connection points and edges denote admissible passenger transfers. Two aggregate streams are considered separately: passengers entering the station building to embark and passengers leaving after disembarking. For each period \(t\),
\[
N_t^{s} \sim \mathrm{Poisson}(\lambda_t^{s}), \qquad s \in \{\mathrm{emb},\mathrm{dis}\}.
\]
At each branching node \(v\), routing proportions are modeled as
\[
\mathbf{p}_{v}^{\,s} \sim \mathrm{Dirichlet}(\boldsymbol{\alpha}_{v}^{\,s}),
\]
and outgoing edge flows follow a multinomial allocation conditional on incoming flow. The total inferred flow on edge \(e\) is
\[
F_{e,t}=F_{e,t}^{\mathrm{emb}}+F_{e,t}^{\mathrm{dis}}.
\]

The role of this model is not to recover detailed pedestrian trajectories, but to propagate uncertainty from aggregate traffic information to the graph edges associated with maintained assets. In practical use, the parameters $\lambda_t^s$ and the Dirichlet hyperparameters $\alpha_v^s$ may be revised whenever updated average traffic information or revised routing assessments become available. This makes the formulation suitable for low-data environments in which only periodic aggregate updates are accessible, rather than continuous direct measurements.

\subsection{Device Load Estimation}

Direct cycle measurements are assumed unavailable. Device loads are therefore estimated from inferred passenger flows using simple stochastic proxy relations.

For door \(d\), with associated passenger flow \(Q_{d,t}\),
\[
L_{d,t}^{\mathrm{door}}=\frac{Q_{d,t}}{\beta_d},
\]
where \(\beta_d\) is a random passenger-per-opening factor centered on the assumption of approximately two passengers per opening.

For elevator \(j\), with assigned passenger flow \(Q_{j,t}\),
\[
L_{j,t}^{\mathrm{elev}}=\frac{Q_{j,t}}{\gamma_j},
\]
where \(\gamma_j\) is an effective occupancy parameter reflecting practical under-utilization of nominal elevator capacity.

\subsection{Maintenance-Condition Probability}

For asset \(i\) and maintenance class \(m \in \{\mathrm{min},\mathrm{med},\mathrm{maj}\}\), let \(A_{i,t}^{(m)}\) and \(C_{i,t}^{(m)}\) denote elapsed age and accumulated cycles since the relevant reset. Thresholds \(T_{i,a}^{(m)}\) and \(T_{i,c}^{(m)}\) are treated as uncertain. The event that maintenance condition has been reached is
\[
\mathcal{M}_{i,t}^{(m)}=
\left\{
A_{i,t}^{(m)} \ge T_{i,a}^{(m)}
\ \text{or}\
C_{i,t}^{(m)} \ge T_{i,c}^{(m)}
\right\},
\]
with probability
\[
P_{i,t}^{(m)}=\mathbb{P}\!\left(\mathcal{M}_{i,t}^{(m)}\mid\mathcal{Y}_{1:t}\right).
\]
Minor service resets only minor counters, medium resets medium and minor counters, and major resets all counters.

This probability-based formulation is important because the proposed framework does not attempt to infer a continuously observed degradation state. Instead, it estimates whether the asset has entered a region in which the corresponding maintenance action becomes advisable. In this sense, the output is not a fault diagnosis signal but an uncertainty-aware maintenance indicator derived from proxy operational usage and threshold variability.

\section{Maintenance Scheduling Model}
\label{sec:scheduling}
The maintenance-condition probabilities obtained in the previous section provide the uncertainty-aware state information used for maintenance planning. In the considered setting, the maintenance categories themselves are predefined by operational practice and service procedures; therefore, the decision problem is not to select an action type freely, but to determine suitable execution times for the corresponding activities. The role of the scheduling model is to align these predefined interventions with estimated asset usage while accounting for service cost, operational disruption, delay penalties, and the potential benefits of grouped interventions within the same asset class.
\subsection{Scheduling Objective}

The maintenance categories are predefined by operational practice; therefore, the scheduling problem concerns execution timing rather than action selection. Let the planning horizon consist of periods \(t=1,\ldots,T\), and let \(x_{i,t}^{(m)} \in \{0,1\}\) indicate whether maintenance category \(m\) is scheduled for asset \(i\) in period \(t\).

The objective combines direct service cost, setup cost for grouped interventions, operational disruption, and penalty for delayed servicing beyond estimated maintenance condition.

Accordingly, the expected cost may be written in the form
\begin{equation}
\min_{\mathbf{x}}
\;
\mathbb{E}
\left[
C^{\mathrm{serv}}(\mathbf{x})
+
C^{\mathrm{setup}}(\mathbf{x})
+
C^{\mathrm{disr}}(\mathbf{x})
+
C^{\mathrm{delay}}(\mathbf{x},\mathbf{P})
\right],
\label{eq:scheduling_objective}
\end{equation}
where \(\mathbf{x}\) denotes the collection of scheduling decisions and \(\mathbf{P}\) denotes the set of maintenance-reaching probabilities.

\subsection{Cost Components}

The direct maintenance execution cost is
\begin{equation}
C^{\mathrm{serv}}(\mathbf{x}) =
\sum_{i}\sum_{m}\sum_{t}
c_{i}^{(m)} x_{i,t}^{(m)},
\label{eq:service_cost}
\end{equation}
where \(c_i^{(m)}\) is the cost of performing maintenance category \(m\) on asset \(i\).

To account for grouped servicing, assets are partitioned into classes such as doors and elevators. Let \(g \in \mathcal{G}\) index asset groups, and let \(z_{g,t}\in\{0,1\}\) indicate whether a maintenance intervention for group \(g\) is opened in period \(t\). Then the corresponding setup cost is
\begin{equation}
C^{\mathrm{setup}}(\mathbf{x}) =
\sum_{g}\sum_{t}
s_g z_{g,t},
\label{eq:setup_cost}
\end{equation}
where \(s_g\) is the fixed setup cost for a group-level intervention. The variables \(z_{g,t}\) are linked to the maintenance decisions \(x_{i,t}^{(m)}\) through standard activation constraints ensuring that \(z_{g,t}=1\) whenever at least one asset in group \(g\) is serviced in period \(t\).

Operational disruption may also depend on both the maintained asset and the maintenance category. A simple additive form is
\begin{equation}
C^{\mathrm{disr}}(\mathbf{x}) =
\sum_{i}\sum_{m}\sum_{t}
d_i^{(m)} x_{i,t}^{(m)},
\label{eq:disruption_cost}
\end{equation}
where \(d_i^{(m)}\) quantifies the inconvenience or operational burden associated with carrying out maintenance of category \(m\) on asset \(i\).

The delay penalty is introduced to discourage prolonged operation after the maintenance condition becomes likely to have been reached. One possible formulation is
\begin{equation}
C^{\mathrm{delay}}(\mathbf{x},\mathbf{P}) =
\sum_{i}\sum_{m}\sum_{t}
\psi_i^{(m)} \, P_{i,t}^{(m)} \, \delta_{i,t}^{(m)},
\label{eq:delay_cost}
\end{equation}
where \(\psi_i^{(m)}\) is a penalty coefficient and \(\delta_{i,t}^{(m)}\) is an indicator that maintenance category \(m\) for asset \(i\) has not yet been executed by time \(t\) despite still being due. In this way, the penalty grows with both the probability of having reached the maintenance condition and the persistence of delayed servicing.

In the present work, the cost coefficients are treated as engineering decision parameters rather than statistically identified quantities. Their role is to encode practical trade-offs between performing maintenance too early, delaying it excessively, and dispatching separate interventions for assets that could be serviced together. This is consistent with the intended use of the framework as decision support in a low-data environment.

\subsection{Grouping Effect and Temporal Alignment}

A practical advantage of the formulation is that it allows less urgent services to be aligned with more urgent ones within the same asset class, reducing setup effort without changing maintenance category. This is relevant in the considered station setting, where only a small number of doors and elevators are maintained and separate dispatch for each asset may be inefficient. The scheduling objective therefore balances delay risk against the organizational benefit of grouped interventions.

\subsection{Reference Calendar Policy}

The proposed schedule is compared with a calendar-based baseline in which each maintenance category is executed at fixed time intervals independently of inferred usage. This baseline ignores asset-specific loading differences and serves as a reference for assessing the benefit of proxy-flow-based scheduling.

\section{Case Study and Simulation Setup}
\label{sec:simulation}

\begin{figure}[!t]
\centering
\includegraphics[width=0.9\columnwidth]{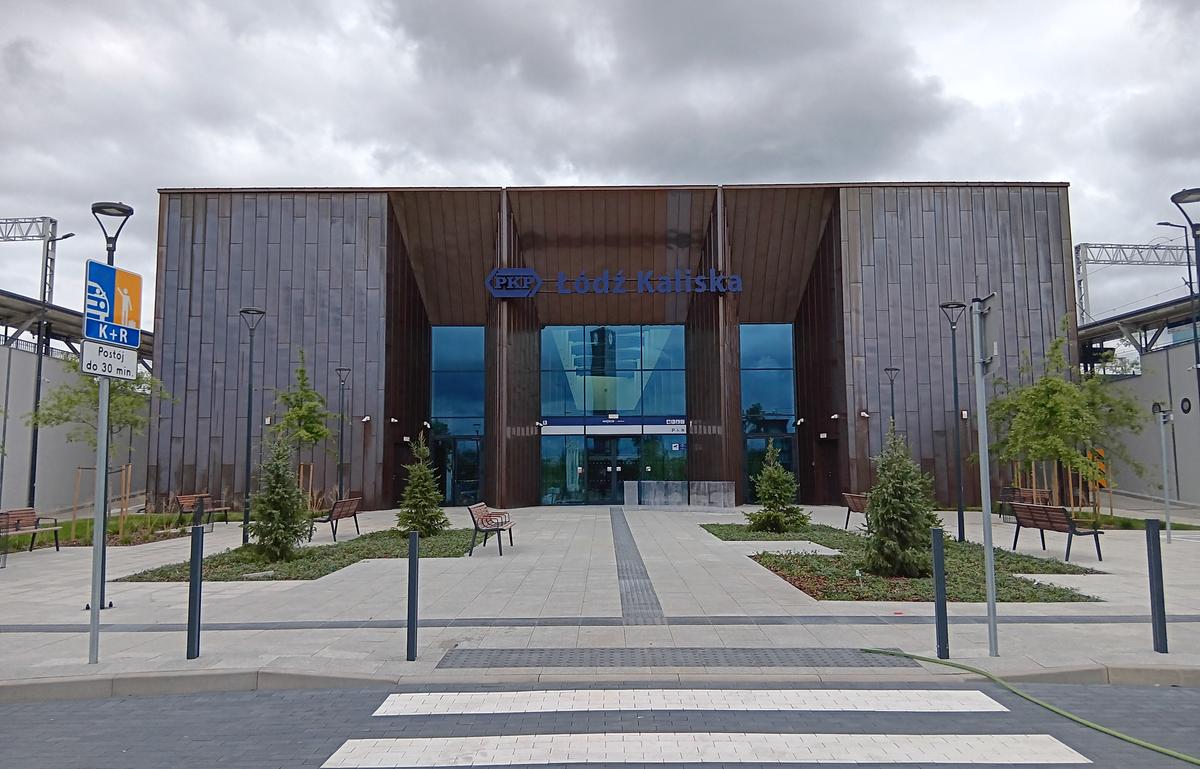}
\caption{View of the recently renovated \L{}\'od\'z Kaliska railway station, which provides the real architectural context for the illustrative case study considered in this paper. The analytical results presented later are based on synthetic numerical data consistent with this layout, since the operational data used in the underlying project are proprietary.}
\label{fig:station_photo}
\end{figure}

The proposed framework is illustrated using a case corresponding to the recently renovated \L{}\'od\'z Kaliska railway station. The station provides a suitable example of a contemporary public-service facility in which passenger movement is structured by a clearly identifiable building layout and by access infrastructure such as automatic doors and elevators. In the present study, the station is represented by a reduced directed graph superimposed on the building plan, with graph nodes corresponding to principal entrances, internal decision points, and access paths relevant to the maintained assets. A station photograph is included in Fig.~\ref{fig:station_photo} in order to emphasize the real architectural context of the considered application.

The case study is motivated by solution-development work carried out within the project financing the present research. In that broader context, operational and maintenance-related information from the station is used in the development of the target analytical solution. However, most of these data are proprietary and cannot be disclosed in a publication. For this reason, the present paper does not attempt to reproduce the real operational dataset. Instead, the proposed method is demonstrated on synthetic numerical data constructed so as to preserve the structure of the actual problem while avoiding disclosure of sensitive information.

The maintained assets considered in the simulation comprise four automatic doors and four elevators linked to selected graph elements. Separate aggregate passenger streams are assumed for embarking and disembarking passengers, with uncertain totals and routing shares consistent with expert observations of station use. Synthetic parameters are used for traffic intensity, routing proportions, load-conversion factors, maintenance thresholds, and cost coefficients. For doors, the passenger-to-cycle conversion is centered on approximately two passengers per opening. For elevators, the cycle estimate is based on uncertain effective occupancy reflecting typical small-group use. Each asset follows a nested maintenance program with minor, medium, and major service thresholds defined in both time and cycles.

The proposed scheduling model is compared with a calendar-based baseline in which maintenance is performed at fixed intervals independently of inferred usage. The objective of the simulation is therefore not to reproduce exact station behavior, but to assess whether proxy-flow-based scheduling can better align interventions and reduce expected maintenance burden under realistic low-data assumptions.

In practical deployment, the proposed framework would require only periodic updates of aggregate passenger-flow estimates, expert revision of selected routing proportions, and maintenance-log information sufficient to update service histories. This makes the approach substantially lighter than sensor-centered predictive maintenance architectures, while still allowing uncertainty-aware maintenance planning for assets whose operating burden is strongly linked to passenger movement through the station.
\section{Results and Discussion}
\label{sec:results}

\subsection{Passenger-Flow Inference on the Station Graph}

\begin{figure}[!t]
\centering
\includegraphics[width=0.9\columnwidth]{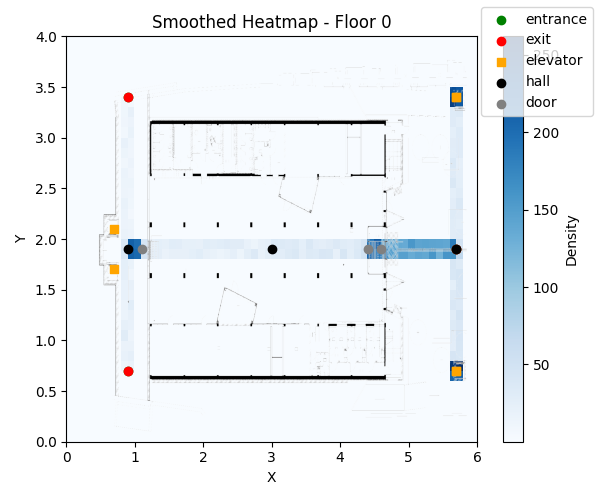}
\caption{Illustrative passenger-flow intensity over the reduced station graph superimposed on the station floor plan.}
\label{fig:graph_heatmap}
\end{figure}

The first result of the proposed framework is the inference of passenger flows over the reduced graph representing the circulation structure of the \L{}\'od\'z Kaliska station building. Fig.~\ref{fig:graph_heatmap} shows the corresponding flow-intensity map projected onto the station layout. As expected, the inferred flows concentrate on the principal access corridors and decision points, while peripheral connections carry visibly smaller traffic volumes. This confirms that even aggregate passenger counts combined with expert-defined routing shares are sufficient to generate differentiated usage patterns over the maintained part of the station.

The reduced graph is constructed by identifying the principal passenger-transfer points visible in the station layout, such as entrances, hall zones, elevator access points, and exits from the maintained part of the building. Graph edges represent feasible aggregate movement directions between these points rather than detailed walking trajectories. This reduction is deliberate: the objective is to retain the connectivity structure relevant to infrastructure loading while avoiding unnecessary geometric detail that would not be supported by the available data.

In the present study, the numerical values of these flows are synthetic, because the operational data used in the broader project context are proprietary. Nevertheless, the flow patterns remain consistent with the real station layout and therefore provide a meaningful basis for testing the proposed maintenance-support logic. The role of this stage is not to reconstruct microscopic pedestrian movement, but to derive an uncertainty-aware operational-demand field from which device loads can be inferred.

\subsection{Estimated Loads of Doors and Elevators}

The inferred graph flows are next translated into approximate operating-cycle loads for automatic doors and elevators. Representative estimates with uncertainty bands are shown in Fig.~\ref{fig:load_bands}. The results clearly separate more heavily used assets from less loaded ones, despite the absence of direct cycle counters or condition-monitoring signals.

For doors, the estimated cycle loads follow passenger flow relatively closely, with uncertainty driven mainly by traffic inference and by the assumed passenger-per-opening coefficient. For elevators, the uncertainty is naturally larger, since the conversion from passenger demand to cycle count is more indirect and depends on effective occupancy. This difference is operationally plausible in the considered station setting and should be viewed as a desirable feature of the framework: instead of forcing a deterministic load estimate, the method makes uncertainty explicit where the proxy relation is weaker.

From the maintenance perspective, this stage is important because it converts passenger movement into quantities directly interpretable as service-related usage. In this way, the reduced station graph acts as an intermediate analytical layer between aggregate traffic information and asset-level maintenance indicators.

\subsection{Maintenance-Condition Reaching Probabilities}

The estimated cycle loads, together with elapsed service time, are propagated into maintenance-reaching probabilities for the nested service categories. Fig.~\ref{fig:maintenance_probabilities} illustrates representative trajectories of these probabilities for selected assets. The results show that assets belonging to the same technical class may approach their maintenance condition at noticeably different rates when their inferred usage differs.

This observation highlights the main weakness of purely calendar-based maintenance in the considered setting. Even when devices share the same nominal maintenance program, their effective operating burden may differ substantially because of station layout and passenger-routing structure. The proposed framework captures this heterogeneity without requiring direct sensing. Moreover, the probabilistic formulation provides a more informative maintenance indicator than a single deterministic threshold-crossing time, because it represents the uncertainty associated with both usage estimation and threshold variability.

\subsection{Grouped Scheduling Versus Calendar-Based Maintenance}

The final stage compares the proposed scheduling framework with the reference calendar policy. An example summary of resulting maintenance plans is given in Table~\ref{tab:schedule_summary}. In the proposed approach, maintenance activities are shifted in time to better reflect inferred asset usage, and selected interventions are aligned within the same asset class whenever this can be achieved without excessive delay penalty.

In the considered station-scale example, the grouping effect is naturally modest in absolute terms because only four doors and four elevators are included. Even so, the comparison indicates two practical benefits. First, lightly loaded assets need not be serviced as early as they would under a purely calendar-based policy. Second, moderately misaligned interventions can be combined into fewer service sessions, reducing setup effort and organizational burden. This is particularly relevant for station infrastructure, where the cost of dispatching maintenance personnel may be significant relative to the simplicity of the serviced device.

Overall, the illustrative results support the main claim of the paper: proxy operational data derived from passenger-flow information can improve maintenance timing relative to fixed schedules, even in a low-data environment without dedicated condition monitoring. The example should not be interpreted as a precise quantitative assessment of the \L{}\'od\'z Kaliska station itself, since the numerical values are artificial. Rather, it demonstrates that the proposed analytical chain remains operationally meaningful when anchored in a real station layout and maintenance context.





\begin{figure}[!t]
\centering
\includegraphics[width=0.9\columnwidth]{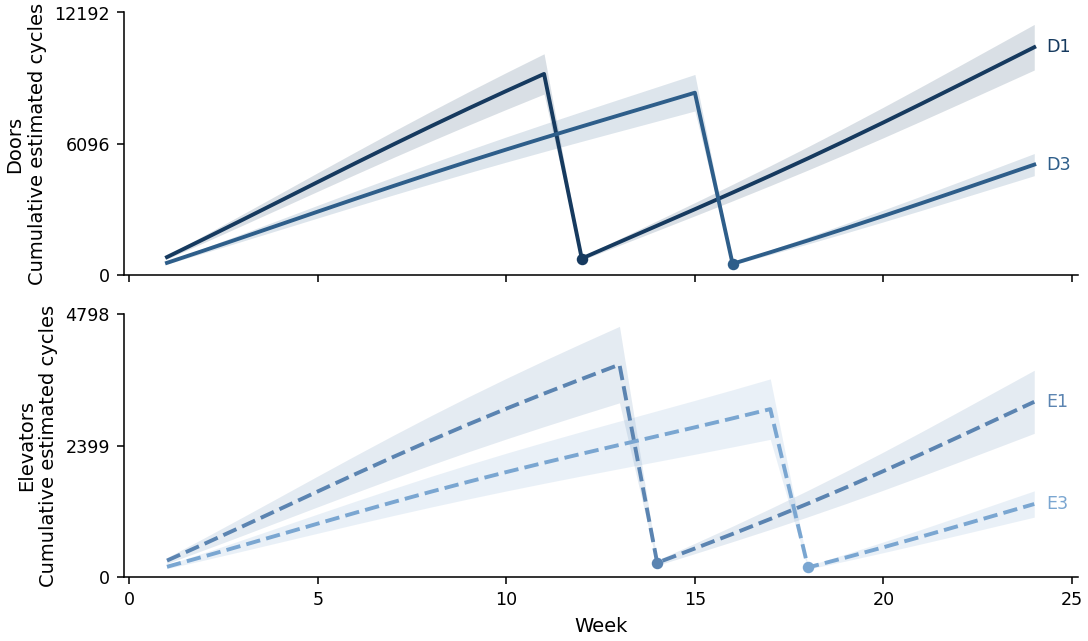}
\caption{Illustrative proxy-based cumulative estimated cycles since last relevant maintenance for selected doors and elevators, shown with uncertainty bands.}
\label{fig:load_bands}
\end{figure}





\begin{figure}[!t]
\centering
\includegraphics[width=0.9\columnwidth]{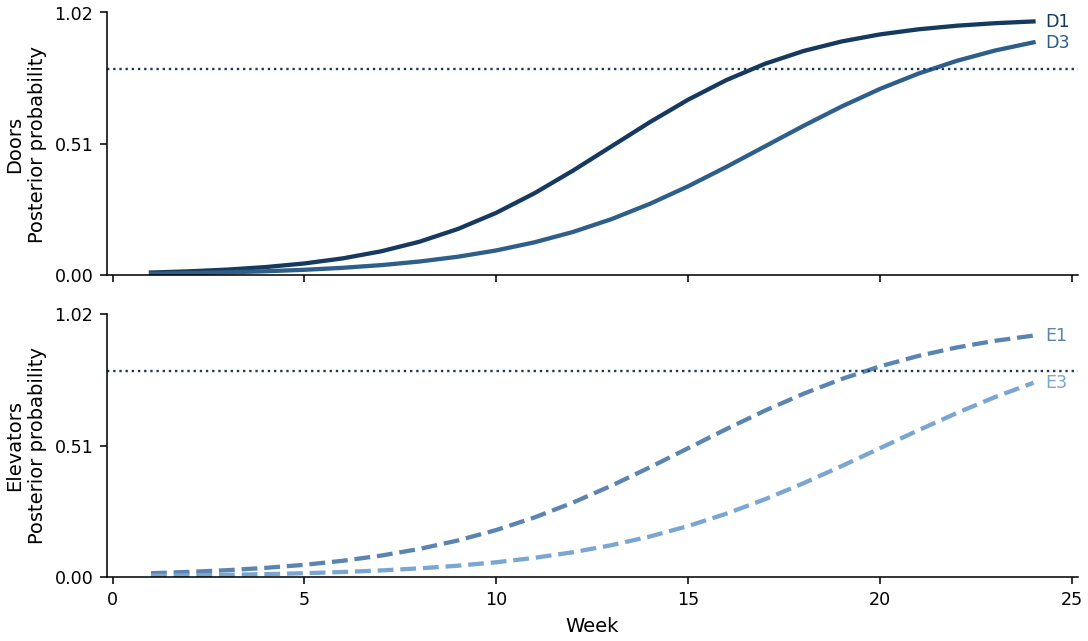}
\caption{Illustrative maintenance-reaching probabilities for selected assets under uncertain cycle and age accumulation.}
\label{fig:maintenance_probabilities}
\end{figure}

Taken together, Figs.~\ref{fig:load_bands} and~\ref{fig:maintenance_probabilities} illustrate the key analytical transition in the proposed framework: uncertain proxy traffic is first converted into cumulative usage estimates and then into maintenance-condition probabilities. This two-stage propagation is important because it preserves the distinction between operational demand and maintenance relevance. A heavily loaded asset does not automatically require immediate intervention, but it approaches its maintenance condition more rapidly and with greater confidence than a lightly loaded one.




\begin{table}[!t]
\caption{Illustrative comparison of maintenance schedules obtained from the calendar-based baseline and from the proposed proxy-flow-based scheduling framework. }
\label{tab:schedule_summary}
\centering
\small
\setlength{\tabcolsep}{0pt}
\begin{tabular*}{\columnwidth}{@{\extracolsep{\fill}}llccc@{}}
\toprule
Asset / policy & Maint. & Cal. & Prop. & Mode \\
\midrule
D1 & M1 & 12 & 11 & G \\
D2 & M1 & 13 & 11 & G \\
D3 & M2 & 15 & 15 & I \\
D4 & M1 & 14 & 11 & G \\
\addlinespace[0.2em]
E1 & M1 & 10 & 10 & G \\
E2 & M1 & 11 & 10 & G \\
E3 & M2 & 16 & 16 & I \\
E4 & M3 & 20 & 20 & I \\
\midrule
\multicolumn{5}{l}{\textbf{Summary}} \\
\cmidrule(lr){1-5}
Calendar policy & -- & \multicolumn{2}{c}{6 sessions} & 0 grouped \\
Proposed policy & -- & \multicolumn{2}{c}{4 sessions} & 4 grouped \\
\bottomrule
\end{tabular*}

\vspace{0.3em}
{\fontsize{7}{8}\selectfont
M1 -- minor maintenance, M2 -- medium maintenance, M3 -- major maintenance; 
Cal. -- calendar-based schedule, Prop. -- proposed schedule; 
Mode: G -- grouped execution, I -- individual execution.}
\end{table}

\section{Conclusion}
\label{sec:conclusion}

A low-data predictive maintenance framework for railway-station doors and elevators was proposed. The method combines Bayesian passenger-flow inference on a reduced station graph with proxy device-load estimation, probabilistic maintenance-condition assessment, and cost-aware grouped scheduling. The simulated case study shows that such a framework can improve maintenance alignment relative to calendar-based policies without requiring additional sensing infrastructure. The main value of the approach lies in using already available proxy operational data to support maintenance planning under uncertainty. Future work will address calibration using maintenance records and integration with broader station digital-twin models. Beyond the specific station example considered here, the same modeling logic may be relevant to other unsensored public-service or building-infrastructure assets whose operating burden can be inferred from aggregate usage proxies rather than direct technical measurements.

Although illustrated for railway-station doors and elevators, the same modeling principle may be applicable to other unsensored public-service assets whose maintenance burden can be inferred from aggregate usage proxies, such as access-control systems, fleet-side service devices, or selected building-infrastructure components.
\bibliography{references}

@article{Jardine2006CBM,
  author    = {Andrew K. S. Jardine and Daming Lin and Dragan Banjevic},
  title     = {A Review on Machinery Diagnostics and Prognostics Implementing Condition-Based Maintenance},
  journal   = {Mechanical Systems and Signal Processing},
  year      = {2006},
  volume    = {20},
  number    = {7},
  pages     = {1483--1510},
  doi       = {10.1016/j.ymssp.2005.09.012}
}

@article{Lei2018Prognostics,
  author    = {Yaguo Lei and Naipeng Li and Liang Guo and Ningbo Li and Tao Yan and Jing Lin},
  title     = {Machinery Health Prognostics: A Systematic Review from Data Acquisition to {RUL} Prediction},
  journal   = {Mechanical Systems and Signal Processing},
  year      = {2018},
  volume    = {104},
  pages     = {799--834},
  doi       = {10.1016/j.ymssp.2017.11.016}
}

@article{Ahmad2012CBMDecision,
  author    = {Rosmaini Ahmad and Shahrul Kamaruddin},
  title     = {A Review of Condition-Based Maintenance Decision-Making},
  journal   = {European Journal of Industrial Engineering},
  year      = {2012},
  volume    = {6},
  number    = {5},
  pages     = {519--541},
  doi       = {10.1504/EJIE.2012.048854}
}

@article{Nunes2023PdMChallenges,
  author    = {Pedro Nunes and Jos{\'e} Santos and Eug{\'e}nio Rocha},
  title     = {Challenges in Predictive Maintenance -- A Review},
  journal   = {CIRP Journal of Manufacturing Science and Technology},
  year      = {2023},
  volume    = {40},
  pages     = {53--67},
  doi       = {10.1016/j.cirpj.2022.11.004}
}

@article{Percy1996BayesianPM,
  author    = {David F. Percy and Khairy A. H. Kobbacy},
  title     = {Preventive Maintenance Modelling: A Bayesian Perspective},
  journal   = {Journal of Quality in Maintenance Engineering},
  year      = {1996},
  volume    = {2},
  number    = {1},
  pages     = {15--24},
  doi       = {10.1108/13552519610113818}
}

@article{Percy1997SparseDataPM,
  author    = {David F. Percy and Khairy A. H. Kobbacy and Bahir B. Fawzi},
  title     = {Setting Preventive Maintenance Schedules When Data Are Sparse},
  journal   = {International Journal of Production Economics},
  year      = {1997},
  volume    = {51},
  number    = {3},
  pages     = {223--234},
  doi       = {10.1016/S0925-5273(97)00054-6}
}

@article{Percy2002BayesianReliability,
  author    = {David F. Percy},
  title     = {Bayesian Enhanced Strategic Decision Making for Reliability},
  journal   = {European Journal of Operational Research},
  year      = {2002},
  volume    = {139},
  number    = {1},
  pages     = {133--145},
  doi       = {10.1016/S0377-2217(01)00177-1}
}

@article{Bousquet2015BayesianGamma,
  author    = {Nicolas Bousquet and Mitra Fouladirad and Antoine Grall and Christian Paroissin},
  title     = {Bayesian Gamma Processes for Optimizing Condition-Based Maintenance under Uncertainty},
  journal   = {Applied Stochastic Models in Business and Industry},
  year      = {2015},
  volume    = {31},
  number    = {3},
  pages     = {360--379},
  doi       = {10.1002/asmb.2076}
}

@article{Haenseler2016TrainStationFacilities,
  author    = {Flurin S. H{\"a}nseler and Michel Bierlaire and Riccardo Scarinci},
  title     = {Assessing the Usage and Level-of-Service of Pedestrian Facilities in Train Stations: A Swiss Case Study},
  journal   = {Transportation Research Part A: Policy and Practice},
  year      = {2016},
  volume    = {89},
  pages     = {106--123},
  doi       = {10.1016/j.tra.2016.05.010}
}

@article{Haenseler2017TrainStationOD,
  author    = {Flurin S. H{\"a}nseler and Nicholas A. Molyneaux and Michel Bierlaire},
  title     = {Estimation of Pedestrian Origin-Destination Demand in Train Stations},
  journal   = {Transportation Science},
  year      = {2017},
  volume    = {51},
  number    = {3},
  pages     = {981--997},
  doi       = {10.1287/trsc.2016.0723}
}

@unpublished{BaranowskiSymmetrySubmitted,
  author       = {Jerzy Baranowski},
  title        = {Asymmetric Relations Between Needs and Data Availability in Predictive Maintenance Systems},
  note         = {Manuscript submitted for publication to \emph{Symmetry}},
  year         = {2026}
}

@article{Gerum2019RailScheduling,
  author    = {Philipp Gerum and Nezir Aydin Altay and Melike Baykal-G{\"u}rsoy},
  title     = {A Data-Driven Predictive Maintenance Scheduling Framework for Localized Railway Track Defects},
  journal   = {Transportation Research Part C: Emerging Technologies},
  year      = {2019},
  volume    = {100},
  pages     = {173--192},
  doi       = {10.1016/j.trc.2019.01.013}
}

@article{Consilvio2024RailPriority,
  author    = {Antonio Consilvio and Roberto D'Auria and Luigi Biggiero and Luca Di Maio},
  title     = {A Decision Support Method for Maintenance Prioritization in Railways Considering Passenger Impact},
  journal   = {European Transport Research Review},
  year      = {2024},
  volume    = {16},
  number    = {1},
  pages     = {1--18},
  doi       = {10.1186/s12544-023-00631-z}
}
\bibliographystyle{IEEEtran}
\end{document}